\documentclass{llncs}

\usepackage{llncsdoc}

\usepackage{xspace}
\usepackage{array}
\usepackage{cleveref}
\usepackage{amssymb,amscd,mathtools}
\usepackage{balance}
\usepackage{paralist}
\usepackage{soul}
\usepackage[usenames,dvipsnames]{xcolor}
\usepackage{tikz}
\usepackage{tabularx}
\usepackage{xspace}
\usepackage{multirow}
\usepackage{cite}
\usepackage[nomessages]{fp}
\usepackage{hyperref}

\usepackage[belowskip=-10pt,aboveskip=0pt]{caption}
\usepackage[caption=false, font=footnotesize]{subfig} 

\usepackage{verbatim}

\newcommand{\protocolName}{PADS\xspace}

\newcommand{\newpar}[1]{\noindent{\bf #1.}}

\newcolumntype{C}[1]{>{\centering\let\newline\\\arraybackslash\hspace{0pt}}m{#1}}

\let\orgautoref\autoref
\renewcommand{\autoref}
{\def\equationautorefname{Eq.}%
	\def\figureautorefname{Fig.}%
	\def\subfigureautorefname{Fig.}%
	\def\Itemautorefname{Item}%
	\def\tableautorefname{Table}%
	\def\algorithmautorefname{Algorithm}%
	\def\paragraphautorefname{Paragraph}%
	\def\sectionautorefname{Section}%
	\def\subsectionautorefname{Section}%
	\def\subsubsectionautorefname{Section}%
	\def\chapterautorefname{Chapter}%
	\def\partautorefname{Part}%
	\def\goalautorefname{Goal}%
	\def\reqautorefname{Req.}%
	\def\adviceautorefname{Rule}%
	\def\parameterautorefname{Param.}%
	\def\definitionautorefname{Definition}%
	\def\theoremautorefname{Theorem}%
	\orgautoref}

\newcommand{\entityStyle}[1]{\mathit{#1}}
\newcommand{\funcStyle}[1]{\mathtt{#1}}

\newcommand{\uuuprv}{\ensuremath{\entityStyle{\mathit{Prv}}}}
\newcommand{\prv}{\uuuprv\xspace}
\newcommand{\prvi}[1]{\ensuremath{\uuuprv_{#1}}\xspace}

\newcommand{\uuuvrf}{\ensuremath{\mathit{Vrf}}}
\newcommand{\vrf}{\uuuvrf\xspace}

\newcommand{\adv}{\ensuremath{\mathit{Adv}}\xspace}

\newcommand{\Netw}{\ensuremath{\mathcal{G}}\xspace}
\newcommand{\Netwt}[1]{\ensuremath{\mathcal{G}_{#1}}\xspace}

\newcommand{\uuuVertexes}{\ensuremath{{V}}}
\newcommand{\Vertexes}{\uuuVertexes\xspace}
\newcommand{\Vertexest}[1]{\ensuremath{\uuuVertexes_{#1}}\xspace}

\newcommand{\EEdges}{\ensuremath{{E}}}
\newcommand{\Edges}{\EEdges\xspace}
\newcommand{\Edgest}[1]{\ensuremath{\EEdges_{#1}}\xspace}

\newcommand{\Neigh}[2]{\ensuremath{N_{#1,#2}}\xspace}
\newcommand{\responsive}[1]{\ensuremath{R_{#1}}\xspace}

\newcommand{\size}{\ensuremath{n}}

\newcommand{\uuuukey}{\ensuremath{k}}
\newcommand{\key}{\uuuukey\xspace}
\newcommand{\keyAtt}{\ensuremath{\uuuukey_{\mathit{att}}}\xspace}
\newcommand{\measurement}{\ensuremath{h}\xspace}

\newcommand{\confign}[1]{\ensuremath{h_{#1}}\xspace}
\newcommand{\configs}{\ensuremath{\mathcal{H}\xspace}}

\newcommand{\attRes}[2]{\ensuremath{r_{#1,#2}}\xspace}
\newcommand{\sattRes}[1]{\ensuremath{r_{#1}}\xspace}

\newcommand{\attValOk}{\ensuremath{\mathtt{10}}\xspace}
\newcommand{\attValBad}{\ensuremath{\mathtt{00}}\xspace}
\newcommand{\attValUnknown}{\ensuremath{\mathtt{11}}\xspace}

\newcommand{\uuuRepresentativity}{\ensuremath{\rho}}
\newcommand{\pRepresentativity}{\uuuRepresentativity\xspace}

\newcommand{\uuuVerifToken}{\ensuremath{\phi}}
\newcommand{\pVerifToken}{\uuuVerifToken\xspace}

\newcommand{\prvBitmask}[2]{\ensuremath{x_{#1}^{#2}}\xspace}

\newcommand{\prvBitmaskEl}[3]{\ensuremath{x_{#1}^{#2}[#3]}\xspace}

\newcommand{\uuuuE}{\ensuremath{\mathcal{E}}}
\newcommand{\Esend}{\ensuremath{\uuuuE_\mathit{send}}\xspace}
\newcommand{\Erecv}{\ensuremath{\uuuuE_\mathit{recv}}\xspace}
\newcommand{\Emac}{\ensuremath{\uuuuE_\mathit{hmac}}\xspace}

\newcommand{\Eatt}{\ensuremath{\uuuuE_\mathit{att}}\xspace}
\newcommand{\Emin}{\ensuremath{\uuuuE_\mathit{min}}\xspace}
\newcommand{\Epads}{\ensuremath{\uuuuE_\mathit{\protocolName}}\xspace}

\newcommand{\uuuuT}{\ensuremath{T}}
\newcommand{\TAtt}{\ensuremath{\uuuuT_\mathit{att}}\xspace}
\newcommand{\dTMax}{\ensuremath{\Delta\uuuuT_\mathit{max}}\xspace}

\newcommand{\timestamp}{\ensuremath{\tilde{\uuuuT}}\xspace}

\newcommand{\timeMax}{\ensuremath{T_\textit{MAX}}\xspace}
\newcommand{\timeTolerance}{\ensuremath{\Delta T}\xspace}

\newcommand{\uuumac}{\ensuremath{\funcStyle{mac}}}

\newcommand{\mac}{\uuumac\xspace}
\newcommand{\macp}[2]{\ensuremath{\uuumac_{#1}(#2)}\xspace}

\newcommand{\uuumacKeygen}{\ensuremath{\funcStyle{mac\_keygen}}}
\newcommand{\macKeygen}{\uuumacKeygen\xspace}

\newcommand{\uuumacVerif}{\ensuremath{\funcStyle{mac\_verif}}}
\newcommand{\macVerif}{\uuumacVerif\xspace}
\newcommand{\macVerifp}[3]{\ensuremath{\uuumacVerif_{#1}(#2,#3)}\xspace}

\newcommand{\signature}{\ensuremath{\sigma}\xspace}

\newcommand{\uuuuseed}{\ensuremath{\nu}}
\newcommand{\seed}{\uuuuseed\xspace}
\newcommand{\seedAtt}{\ensuremath{\uuuuseed_{att}}\xspace}

\newcommand{\uuuprngGen}{\ensuremath{\funcStyle{prng\_gen}}}
\newcommand{\prngGen}{\uuuprngGen\xspace}

\newcommand{\prngState}{\ensuremath{\mu}\xspace}
\newcommand{\prngValue}[1]{{\ensuremath{\theta_{#1}}\xspace}}

\newcommand{\uuuSelfAttest}{\ensuremath{\funcStyle{selfAtt}}}
\newcommand{\fSelfAttest}{\uuuSelfAttest\xspace}

\newcommand{\uuuConsensus}{\ensuremath{\funcStyle{cons}}}
\newcommand{\fConsensus}{\uuuConsensus\xspace}

\newcommand{\uuufcheck}{\ensuremath{\funcStyle{check}}}
\newcommand{\fCheck}{\uuufcheck\xspace}

\newcommand{\uuugetSoftConf}{\ensuremath{\funcStyle{getSoftConf}}}
\newcommand{\fGetSoftConf}{\uuugetSoftConf\xspace}

\newcommand{\uuuschedule}{\ensuremath{\funcStyle{schedule}}}
\newcommand{\fSchedule}{\uuuschedule\xspace}

\newcommand{\prob}[1]{\ensuremath{\mathit{Pr}(#1)}\xspace}

\newcommand{\MODIFIED}[1]{{#1}}

\newcommand{\coverage}[1]{\ensuremath{c_{#1}^{t}}\xspace}

\begin{document}

\newcommand{\paperTitle}{\protocolName: Practical Attestation for Highly Dynamic Swarm Topologies}

\title{\protocolName: Practical Attestation for Highly \\Dynamic Swarm Topologies}

\pagestyle{headings}
\titlerunning{\paperTitle}

\author{Moreno Ambrosin\inst{1} \and Mauro Conti\inst{2} \and Riccardo Lazzeretti\inst{3} \and \\  Md Masoom Rabbani\inst{2} \and Silvio Ranise\inst{4}}

\institute{
Intel Corporation, Hillsboro, USA, \email{moreno.ambrosin@intel.com} 
 \and
University of Padua, Padua, Italy, \email{\{conti,rabbani\}@math.unipd.it} 
\and
Sapienza University of Rome, Rome, Italy, \email{lazzeretti@diag.uniroma1.it}
\and
Fondazione Bruno Kessler, Trento, Italy \email{ranise@fbk.eu}}


\maketitle



\begin{abstract}


Remote attestation protocols are widely used to detect device configuration (e.g., software and/or data) compromise in Internet of Things (IoT) scenarios. Unfortunately, the performances of such protocols are unsatisfactory when dealing with thousands of smart devices. 
Recently, researchers are focusing on addressing this limitation. The approach is to run attestation in a collective way, with the goal of reducing computation and communication.
Despite these advances, current solutions for attestation are still unsatisfactory because of their complex management and strict assumptions concerning the topology (e.g., being time invariant or maintaining a fixed topology). 
%
In this paper, we propose PADS, a secure, 
efficient, and practical protocol for attesting potentially large networks of smart devices with unstructured or dynamic topologies. PADS builds upon the recent concept of non-interactive
attestation, by reducing the collective attestation problem into a minimum
consensus one. We compare PADS with a state-of-the art collective attestation protocol and validate it by using realistic
simulations that show practicality and efficiency. The results  confirm the suitability of PADS for low-end devices, and highly unstructured networks.\footnote{\textbf{The paper has been submitted to ESORICS 2018}}

\end{abstract}



    


\sloppy     

\section{Introduction}



Smart Internet of Things (IoT) devices are used in many different fields, ranging from simple small scale systems, such as home
automation, to large scale safety-critical environments, e.g., in military
systems, drone-based surveillance systems, factory automation, or smart
metering. 
Unfortunately, their role in critical systems, their low cost
nature, and their wide usage, make IoT devices an attractive target for
attackers~\cite{symantec_iot_ddos}.  Security in IoT is thus a major concern, to guarantee both the
correct operational capabilities of devices, and prevent data thefts and/or
privacy violations.

Malwares represent a major
threat to IoT systems, through which attackers replace the original firmware 
from the
devices with malicious code, to perform larger
attacks~\cite{hvac_hack,jeep_hack}.
One effective way to detect these types of attacks is remote attestation, an
interactive protocol between a prover (\prv) and a remote verifier (\vrf) that allows \vrf to assess the integrity of \prv's configuration  (e.g., firmware and/or data). In a remote attestation protocol, \vrf sends a challenge to \prv; \prv computes a {\em measurement}
(typically a hash) of its configuration, and returns
such measurement to \vrf, integrity and authenticity-protected with a Message
Authentication Code (MAC), or digitally signed. \vrf then checks whether: (i)
the received data is authentic; and (ii) if the received measurement conforms
to an expected ``good configuration'' (e.g., taken from a database of known
``good configurations'')~\cite{francillon2012systematic,francillon2014minimalist}.  As reported in recent alternative solutions, the device itself can check whether
the configuration is correct, and simply report a (signed or MAC-ed) binary
value indicating whether attestation was successful or not~\cite{ambrosin2016sana}. Furthermore, to better match autonomous systems requirements, attestation can be started by \prv rather than \vrf, 
at predefined points in time~\cite{ibrahim2017seed}. 
The security of remote attestation protocols, in practical settings, is guaranteed by a hybrid combination of software and hardware~\cite{eldefrawy2012smart,brasser2015tytan}, which acts as a root of trust for measurement in the device.

Unfortunately, remote attestation is hard to scale in its basic form. Indeed, its overhead is linear in the number of provers in a system, making it potentially unpractical for very large deployments, e.g., networks of autonomous
and/or collaborating low-end devices. For this reason, in the past years researchers
tried to design novel protocols, called {\em collective attestation
	protocols}, to allow attestation to scale, while retaining useful security
properties~\cite{asokan2015seda,ambrosin2016sana,carpent2017lightweight,kohnhauser2017scapi}. Previous work, however, focused on networks with mostly static topologies, scaling attestation over spanning trees, which are difficult and expensive to maintain in networks with faulty links, and/or dynamic topologies. 
Examples of this type of settings are decentralized coordination of Unmanned Aerial Vehicles (drones)
\cite{valavanis2014handbook,dasgupta2008multiagent} or adaptive driverless
cars traffic flow management \cite{waldrop2015no,fagnant2015preparing,gora2016traffic}. 

\newpar{Idea and Contribution} In this paper, we propose a novel protocol for
efficiently and effectively collect attestation proofs from provers in highly
dynamic networks of autonomous devices.  Inspired by the sensor fusion
literature~\cite{boyd2006randomized}, we take a different approach
with respect to previous works.  
We start from the concept of non-interactive
attestation~\cite{ibrahim2017seed}, and turn the problem of
efficiently collecting attestation proofs from provers, into a distributed
consensus problem: A network of agents (i.e., provers) attest themselves (as in~\cite{ibrahim2017seed}), and then corroborate their individual results
into a ``fused'' attestation result via minimum consensus. The final
collective result will carry sufficient information to tell which devices in
the network are in an {\em healthy} state, i.e., run a correct version of the
firmware, and which are compromised.
Our solution perfectly fits with networks of autonomous devices
that operate without a central unit coordinating the operation.

We achieve this goal through \protocolName, the first protocol that supports
attestation on highly dynamic and unstructured networks. 
\protocolName presents several key advantages w.r.t. existing literature: 

\begin{compactitem}
	
	\item It does not require the establishment of an overlay spanning tree to efficiently carry out the collective attestation process, and as such, it is suitable for unstructured networks;
	
	\item It does not require provers to be always online and/or reachable during the whole attestation process; 
	
	\item It is computationally efficient and suitable for resource-constrained devices, and protocols; 
	
	\item As in~\cite{ibrahim2017seed}, \protocolName starts at (and produces an attestation result related to) given points in time, and the verifier can at any time obtain the status of the network by querying any prover in the network;
	
	\item It opens for intelligent uses of attestation results, e.g., healthy nodes can know the status of other nodes and exclude compromised nodes from computation.
	
\end{compactitem}

We show the performance of \protocolName through simulations, and compare it
against SANA~\cite{ambrosin2016sana}. Our experimental results confirm the
suitability of \protocolName for low-end devices, and highly unstructured
networks.

\section{Related Work} 
Asokan
et al.~\cite{asokan2015seda} first highlighted the challenges in remote
attestation for large swarms of low-end devices, and proposed SEDA, a scalable
protocol for collective attestation. SEDA allows \vrf to perform
attestation over an (overlay) spanning tree, rooted at \vrf; each device
attests its neighbors, and reports back to its parent (in the spanning tree).
Each device in SEDA is equipped with the minimal hardware requirements
necessary for attestation on embedded devices, i.e., a Read-Only Memory (ROM),
and a Memory Protection Unit (MPU)~\cite{francillon2012systematic}. SEDA
provides an efficient mechanism to perform attestation, but requires full
connectivity among nodes during the whole attestation process. DARPA~\cite{ibrahim2016darpa} improves the resiliency of
SEDA against strong attackers, by allowing nodes to collaboratively detect
hardware-compromised devices. Unfortunately, DARPA inherits most of the main
limitations of SEDA when it comes to dynamic networks. Ambrosin et
al.~\cite{ambrosin2016sana} proposed SANA, an end-to-end secure protocol for
collective attestation that, compared to SEDA, limits the strength of hardware
attacks, is publicly verifiable, and does not require all the nodes in the
system to be equipped with a Trusted Execution Environment (TEE), making it
usable in heterogeneous deployments. SANA allows untrusted nodes to aggregate
attestation proofs collected from provers, using a generalized aggregate
multi-signature scheme. While resolving most of the major shortcomings of
SEDA, SANA still requires full connectivity among devices; moreover, it
introduces severe overhead on low-end provers. More
recently,~\cite{carpent2017lightweight} propose two remote attestation
protocols that improve SEDA with respect to scalability and resiliency to
hardware attacks. However,\ \cite{carpent2017lightweight} requires full
connectivity among devices for the whole attestation process. Remarkably, the
recent work in~\cite{kohnhauser2017scapi} improves SEDA by supporting
more dynamic networks, at the price, however of additional complexity in the
system (election of clustered nodes, etc.). More recently, Ibrahim et al. proposed SeED~\cite{ibrahim2017seed}, a protocol that allows to
perform a prover-initiated series of attestation step at different random
points in time. This represents a good fit for several applications, and in
dynamic networks. Unfortunately, SeED leverages SEDA to scale to collective attestation, and thus inherits its limitations.  More recently, Carpent et al.\cite{DBLP:ERASMUS} proposed ERASMUS, a remote attestation protocol for unattended devices which unlike other RA schemes promises to identify mobile adversaries between two successive attestation periods. Although it substantially minimizes prover's computation but \cite{DBLP:ERASMUS} leverages SEDA  or LISA \cite{asokan2015seda,carpent2017lightweight} for collective attestation of swarms. Thus inherits their shortcomings.

\section{System Model and Assumptions}\label{sec:system_model}


\subsection{System Model} \label{sec:ref_system_model}

We consider a network \Netw modeled as a graph with vertices \Vertexes and edges \Edges of interconnected (possibly low-power) devices. 
The number of edges in \Netw is static, with $|\Vertexes| = \size$; however, the topology of the network is highly dynamic, i.e., \Edges may vary over time. 
To make this explicit, we denote with $\Edgest{t}$ the set of edges in $\Edges$ at time instant $t$.  In a similar way, we use subscripts from a set of time instants with all other variables that range over time.
We further indicate with \Neigh{i}{t} the set of all the vertex neighbors of a vertex $i$, at time $t$, i.e., $\Neigh{i}{t} = \{j~\in~\Vertexes~|~\exists~e~\in~\Edgest{t}\text{ s.t., }e\text{ connects }i\text{ and }j\}$.
From a communication standpoint, a device (i.e., a vertex) could be either \emph{unresponsive}, or \emph{inactive}. An unresponsive device is a device with which communications have been interrupted for some reason, for example, intentionally or for physical causes, e.g., the connection experienced interference. An inactive device is a device which does not participate in any communication with other devices, e.g., an isolated device in the network.
A device is active or responsive if it is not inactive and we denote with $\responsive{t}$ the set of devices that are responsive at time instant $t$.

\begin{figure}[ht]
	\centering
	\includegraphics[width=.8\columnwidth]{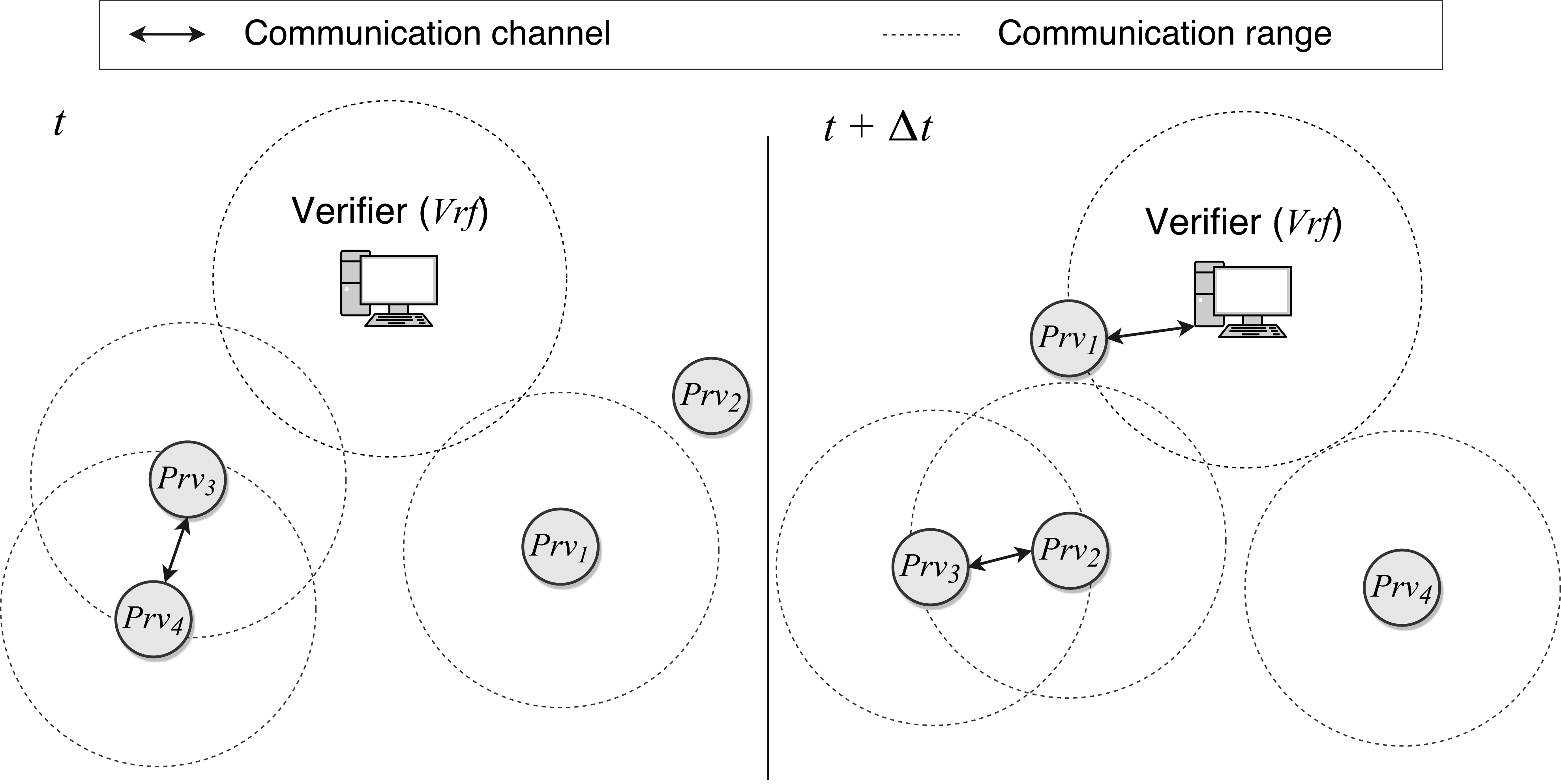}
	
	\caption{Target system model for \protocolName. Provers communicate wirelessly with entities withing their wireless coverage, and are mobile; as such, at different points in time, the topology may change ($t$ and $t+\Delta t$ in the picture). Similarly, the verifier can communicate with any prover in its wireless coverage.}
	\label{fig:system_model}
\end{figure}

Similarly to previous works in this area~\cite{asokan2015seda,ambrosin2016sana}, we define a collective attestation protocol as a protocol between the following entities: prover (\prv), and verifier (\vrf). Each vertex in \Netw is a prover \prvi{i}, and thus, in the remaining of the paper, we will use \prvi{i} to refer to vertex $i$ in \Netw. 
We assume the existence of at least one verifier \vrf, which is not part of \Netw. 
As shown in \autoref{fig:system_model}, \vrf can communicate  only with the provers that are in its 
communication range, as well as responsive, at a certain point in time $t$.

\subsection{Security Model}\label{sec:security_assumptions}

\newpar{Prover Capability Requirements and Assumptions} In line with all the previous works in swarm attestation \cite{asokan2015seda, ibrahim2016darpa, ambrosin2016sana, carpent2017lightweight}, and in particular according to very recent works on non-interactive attestation~\cite{ibrahim2017seed}, we assume each device presents the following minimal features: (i) A Read-Only Memory (ROM), where integrity-protected attestation code should reside; (ii) A Memory Protection Unit (MPU), that allows to enforce access control on areas of the memory, e.g., read-only access to certain memory areas exclusively to attestation protocol code~\cite{francillon2012systematic}; (iii) A secure Real-Time Clock (RTC)~\cite{ibrahim2016darpa}, to tie the generation of the attestation proof to the current time; and (iv) A secure attestation trigger (AT)~\cite{ibrahim2017seed}, i.e., a dedicated circuit that should be non-interruptible by the operating system running on the device.

Existing popular designs for embedded devices, such as SMART~\cite{eldefrawy2012smart}, TrustLite~\cite{koeberl2014trustlite}, or
TyTAN~\cite{brasser2015tytan}, natively provide ROM and MPU capabilities, and
have been used in most (if not all) previous works as reference architectures. The latter two features, RTC and AT, are provided by additional (yet minimal) dedicated hardware components, as shown in~\cite{ibrahim2017seed}.

\newpar{Device Classification} From a security standpoint, in line with
previous works, we classify a device with a provably correct (i.e.,
attestable) configuration as \emph{healthy}; in case the configuration is not
correct, we classify the device as \emph{compromised}.  

However, as we deal with very dynamic networks, i.e., where \Netw over time may be a disconnected graph, \vrf could not be able to always determine the status of a device as healthy or compromised.
On one hand, while healthy nodes usually are assumed to correctly follow the
attestation protocol, it may happen that they do not reply to communication
requests from other devices because inactive, e.g., they are busy, or decide to
turn into an idle state to save battery during inactivity, or unresponsive for
physical causes (e.g., a prover device moving out of range during a
communication). On the other hand, a compromised prover may refuse to respond
to try to evade detection.
As a consequence, considering an unresponsive or inactive prover as
compromised is, in principle, a wrong conclusion.
From a communication perspective, we refer to provers that cannot be reached
over the whole duration of the attestation protocol as {\em unreachable},
while we refer to provers that at least at one point in time take part in the
protocol as {\em reachable}.

Given the above, from a security perspective we consider three
possible outcomes of the attestation process, for a prover: {\bf Healthy}, which means that the device is reachable, and has a correct configuration; {\bf Compromised}, which means that device is reachable, but the running software configuration is incorrect; {\bf Unknown}, which means that the device was unreachable, and thus not ``covered'' by the attestation protocol.









\newcommand{\advSoft}{\ensuremath{\adv_\mathit{soft}}\xspace}
\newcommand{\advComm}{\ensuremath{\adv_\mathit{com}}\xspace}
\newcommand{\advMob}{\ensuremath{\adv_\mathit{mob}}\xspace}

\newpar{Adversary Model} We consider the following types of adversary:

\begin{compactitem}
	
	\item {\em Communication Adversary (\advComm):} Conforms to the Dolev-Yao model~\cite{dolev1983security}. \advComm is assumed to be fully in control of the communication channel between provers, and between provers and \vrf. This adversary can perform several types of attacks: (i) drop attestation reports from provers; (ii) replay attestation reports from provers, e.g., replay correct attestation reports even after a device is compromised; (iii) forge attestation reports from provers to change their value.
	
	\item {\em Software Adversary (\advSoft):} A purely software adversary can exploit vulnerabilities in provers' software, e.g., to remotely inject malware (i.e., modify existing code, introduce new malicious code, or read unprotected memory locations). \advSoft may adopt several strategies to remain undetected, such as: (i) Modify the code responsible for attestation inside the prover; (ii) Extract \prv's security parameters; and (iii) Manipulate \prv's clock to trigger future correct measurements of current software~\cite{ibrahim2017seed}.
	
	\item {\em Mobile Software Adversary (\advMob):} It is a software adversary that tries to evade detection by ``physically moving''. As an example, compromised provers can keep sufficient distance from other provers to remain ``undetected'' by other approaching devices.
	
\end{compactitem}


Finally, similarly to previous work in the area, we consider both physical adversaries and DoS attacks on provers out of the scope of this work.



\newpar{Key Management} Key management has been studied extensively over the past years in several fields, e.g., in the context of Wireless Sensor Networks and Internet of Things (the reader may refer to~\cite{zhou2008securing} for a comprehensive survey of main approaches). While several previously proposed schemes may be adopted in our design, for the sake of simplicity we consider two main options:  (1) enable the use of a na\"ive
master key \keyAtt shared by all the $n$ devices in \Netw, and kept in a hardware-protected key storage, similar
to~\cite{carpent2017lightweight}; or (2) adopt individual public/private key pair (and certificate) for each device, in order to allow message authentication or to exchange pair-wise symmetric keys. The choice here, as suggested in \cite{carpent2017lightweight}, depends on the type of adversary from which the scheme wants to protect against: indeed, it is clear that a shared master key would not mitigate an hardware attacker, while the use of (2) would limit the impact of an hardware compromise, at the price, however, of an higher cost in terms of performance.\footnote{ECDSA signature generation can take a thousand time more compared to a SHA-1 based MAC, to be computed on a resource-constrained platform~\cite{asokan2015seda}.} Both options are valid, and the
decision should be made depending on the trade-off between performance and
security. In what follows, we target software-only attackers and, for ease of exposition, similarly to~\cite{carpent2017lightweight}, we consider the
symmetric case (1) to describe and evaluate \protocolName. Appendix \ref{sec:summary_and_discussion} briefly discusses potential consequences of having a stronger adversary in the system.









\newpar{Requirements}
We consider the following requirements for a secure non-interactive collective attestation in a highly dynamic environment:

\begin{compactitem}
	
	\item {\em Resiliency}. The protocol must be resilient to cases where nodes mobility causes failures in (and consequent modifications to) the communication links between devices.
	
	\item {\em Efficiency}. The lack of a topology makes collective attestation harder; nevertheless, the protocol should be efficient for large scale network of resource-constrained devices.

	\item {\em Heterogeneity}. The collective attestation protocol shall support an heterogeneous population of devices, with different configurations.
	
	\item {\em Unforgeability}. The collective attestation protocol shall guarantee that no software compromised device can forge an attestation result.
	
	\item {\em Low complexity}. The collective attestation protocol should not require complex network and/or routing management.
	
\end{compactitem}

\section{Preliminaries and Notation}
\label{sec:preliminaries_and_notation}

\subsection{Minimum consensus}\label{sec:min_consensus}

In this paper we rely on the {\em minimum consensus} protocol \cite{olfati2005consensus} to share attestation results among provers, and iteratively converge to a ``snapshot'' of the status of the network. \vrf can obtain the collective status of the network by query {\em any} prover in \Netw.
Consensus algorithms have been
proposed to overcome the necessity  of distributed and fault-tolerant
computation and information exchange algorithms. 
These protocols perfectly fit
the constrains of networks characterized by: (i) no centralized entity
coordinating computation, communication and time-synchronization, (ii)
topology not completely known to the nodes of the network, and (iii)  limited
computational power and energy resources \cite{boyd2006randomized}. Consensus
algorithms have wide application in wireless sensor networks
\cite{olfati2004consensus,olfati2005consensus}, unmanned air vehicles (UAVs)
coordination \cite{ren2005survey}, swarm flocking
\cite{blondel2005convergence}, and many other application scenarios
attributable to IoT.

Of particular interest for our work are asynchronous consensus protocols
~\cite{olfati2004consensus,olfati2005consensus}, where nodes
periodically broadcast their status to neighbors within their connectivity
radius. In practice, any node receives information by the neighborhood, updates his status\footnote{There is no need for a real update synchronization among the nodes.} and shares the new state to his neighbors.  By having each node iterating such step multiple times, the state of
all the network nodes converges to the same consensus.
The minimum consensus protocol \cite{olfati2005consensus} we are using in \protocolName distributively computes the minimum of the
observations (attestation) provided by the nodes of the network. 
The input sequence of the minimum consensus protocol is represented by
the sequence \prvBitmask{}{0}=~(\prvBitmask{1}{0},~\prvBitmask{2}{0},~\ldots,~
\prvBitmask{n}{0}), representing the inputs of the $n$ nodes in the network.  According to
\cite{olfati2004consensus}, given \prvBitmask{}{0}, the
protocol generates a sequence of observation states
$\{\prvBitmask{}{t}\}_{t=0}^{\inf}$ and in the step $t$ any node $i$ of the
network updates its status by computing $\prvBitmask{i}{t+1} = min_{j\in N_i
	\cup \{i\}} \prvBitmask{j}{t}$, where $N_i$  is the set of neighbors of the
node $i$  (more details are provided in
\autoref{sec:protocol_description}).
%

Finally, previous research works~\cite{yadav2007distributed,dimakis2010gossip} showed that distributed consensus protocols converge in a finite amount of time (i.e., steps), whose upper bound depends on the structure of the graph. 

\subsection{Notation}\label{sec:notation}


\autoref{tbl:notation_table} introduces the notation we will use in this paper. Let $\{0,1\}^\ell$ represent the set of all bit strings of length $\ell$. We use $\gets_R~M$ to represent a uniformly random sampling from a set $M$, and with $|M|$ the cardinality of a set $M$.
\protocolName security relies on {\em message authentication code} (MAC), and {\em pseudo-random number generator} (PRNG).

A MAC is a tuple of probabilistic polynomial time algorithms (\macKeygen, \mac, \macVerif). \macKeygen takes as input a security parameter $1^{\ell_\uuumac}$, and outputs a symmetric key \key; $\signature \gets \macp{\key}{m}$ and $\macVerifp{\key}{\signature}{m} \in \{1,0\}$ are, respectively, the computation of a MAC on a message $m$ using a key \key, and the verification of the MAC \signature computed on a message $m$ using key \key (1 = valid MAC, 0 otherwise).

A PRNG is a probabilistic polynomial time algorithm \prngGen. \prngGen takes as input the previous state $\prngState_{t-1}$ (at time $t-1$), and outputs a \mbox{(pseudo-)}random value $\prngValue{t} \in \{0,1\}^{\ell_\prngValue{}}$, as well as its current state $\prngState_{t}$.

\newcommand{\tblSpace}{\\[.0cm]}

\begin{table}[!t]
	\centering
	\footnotesize
	\def\arraystretch{1.1}
	\caption{Notation.}\label{tbl:notation_table}
	\begin{tabularx}{\columnwidth}{p{2.5cm} X}

		\hline 
		
		\multicolumn{2}{c}{\bf \small Entities}
		\\
		\hline
		
		\Netwt{t} = (\Edgest{t}, \Vertexest{t}) & Network of provers at time $t$, with edges \Edgest{t} and vertices \Vertexest{t}.
		\tblSpace

		\prv, \vrf      & Prover and verifier, respectively.
		\tblSpace
		
		\Neigh{i}{t}    & Neighbors of prover $\prvi{i}$ at time $t$, in \Netwt{t}.
		\tblSpace
		
		\responsive{t}  & Set of active and responsive provers at time $t$. 
		\tblSpace
		
		\hline
		\multicolumn{2}{c}{\bf \small Parameters}
		\\
		\hline

		\size               & Size of the network, i.e., number of provers in \Netw.
		\tblSpace
		
		\keyAtt             & Symmetric attestation key shared among provers in \Vertexes.
		\tblSpace
		
		\TAtt               & Time at which an attestation is performed.
		\tblSpace
		
		\dTMax              & Maximum time interval between two consecutive attestation runs.
		\tblSpace
		
		\seed               & PRNG secret seed, shared among provers.
		\tblSpace
		
		\attRes{i}{t}       & Attestation result of \prvi{i} at time $t$.
		\tblSpace
		
		\prvBitmask{i}{t}   & Observation of network's (attestation) status from \prvi{i} at time $t$, in the form of a bitmask of $2\times\size$ bits.
		\tblSpace
		
		\prvBitmaskEl{i}{t}{j}   & $j$-th bits couple in \prvBitmask{i}{t} representing the status of $j$-th prover in \Netw. 
		\\
		
		
		






		\hline

	\end{tabularx}
	
\end{table}

\section{\protocolName: Efficient Attestation}\label{sec:protocol_description}


\subsection{Protocol Rationale and Overview}\label{sec:protocol_overview}

The goal of \protocolName is to cope with settings with a high mobility
degree. To this end, \protocolName achieves network attestation by
corroborating individual attestation reports from devices, via consensus. 

At a high-level, \protocolName works as follows. At the same point in time
\TAtt, each prover \prvi{i} performs a local self-attestation step, checking
whether its software configuration (i.e., its measurement \measurement)
corresponds to a known ``good configuration''; this is achieved by matching it
against a list of pre-installed, and thus static, configurations \configs~(similar
to~\cite{ambrosin2016sana}). If the configuration is ``good'', i.e.,
$\measurement \in \configs$, \prvi{i}'s local attestation procedure outputs
$\sattRes{i}=1$, and $\sattRes{i}=0$ otherwise. 

Minimum consensus is used to spread the knowledge about each node state through the network. In order to let \vrf to obtain a ``snapshot'' of the status of the network from {\em any} prover, we let each prover maintain the status of the whole network, and update it iteratively.
To ease an hardware implementation (see \autoref{sec:protocol_details} for the details), we represent the three attestation outcomes introduced in \autoref{sec:security_assumptions} using the following 2 bits values:

\begin{compactitem}
	
	\item {\bf Compromised} (\attValBad): responsive compromised prover.
	\item {\bf Healthy} (\attValOk): responsive healthy prover.
	\item {\bf Unknown} (\attValUnknown): healthy or compromised prover that is unreachable.
	
\end{compactitem}

It follows that, from a consensus perspective, an observation for a prover \prvi{i} \prvBitmask{i}{t} is a bitmask of $2\times\size$ bits, reporting attestation information for the whole network \Netw. We indicate with \prvBitmaskEl{i}{t}{j}, the attestation information relative to prover $j$ in \Netw, i.e., the $j$-th pair of bits in \prvBitmask{i}{t}.

When the consensus protocol starts after attestation, a prover \prvi{i} has only knowledge about its attestation outcome \sattRes{i}, and thus sets $\prvBitmaskEl{i}{0}{i} =\sattRes{i}\in \{\attValBad,~\attValOk\}$; having no knowledge on the state of other devices yet, the remaining pairs of bits in \prvBitmask{i}{0} are set to \attValUnknown, i.e., to represent the ``unknown'' state.
In subsequent steps, provers exchange and combine their observations through the consensus algorithm, this way iteratively building a unique ``view'' of the status of the network. 
During a generic step $t$ any node $i$ broadcasts its (MAC-ed) observation \prvBitmask{i}{t} about the network and receives (MAC-ed) observations \prvBitmask{j}{t} from active devices $j\in \Neigh{i}{t}$\footnote{Synchronization among devices is not really needed in the protocol and has been introduced for simplicity.}. Note that, a compromised device transmitting a fake message can be identified from a wrong MAC (recall that, as stated in \autoref{sec:security_assumptions}, the MPU in the device can prevent access to the memory area where \key resided); 
$\min_\mathit{att}$ combines the observations of \prvi{i} and the received observations from \prvi{j} as:
\begin{align}\label{eq:min_consensus_att}
	&\prvBitmask{i}{t+1} = {\textstyle \min_\mathit{att}}(\prvBitmask{i}{t},\{\prvBitmask{j}{t}\}_{\prvi{j}\in \Neigh{i}{t}\cap \responsive{t}}) = x;\\\nonumber
	& x,~s.t.,~x[l] = \min(\prvBitmaskEl{i}{t}{l},\{\prvBitmaskEl{j}{t}{l}\}_{\prvi{j}\in \Neigh{i}{t}\cap \responsive{t}}),~l = 0,...,\size.
\end{align}
Similarly other nodes update their status.
%
Note that, if we restrict the set of possible values for every \prvBitmaskEl{i}{t}{j} to $\{\attValBad,~\attValOk,~\attValUnknown\}$, \autoref{eq:min_consensus_att} can be efficiently computed using the $\mathtt{AND}$ operator.

In order to guarantee the correctness of the consensus results, the
value of \prvBitmask{i}{t} is stored in a access-restricted area of \prvi{i},
protected by MPU inside the device; furthermore, messages are integrity protected by a MAC computed using a
common key \keyAtt shared by all devices, and secured in an read/write protected area of the memory. Repeating updates using the consensus algorithm through all the network spreads this information to all nodes in \Netw.

%

Finally, in order to obtain the status of the network, \vrf queries any
device \prvi{i} in \Netw; if the latter is not compromised (\vrf verifies this by asking \attRes{i}{t} from \prvi{i}), it will return the consensus state representing its
knowledge about each node.  This allows \vrf to obtain an immediate
approximation of the status of the network, limited by some uncertainty; such uncertainty derives from provers not replying, and/or information that has not yet reached the node. We underline that \vrf can query a \prv anytime during the protocol. However closer is the query to \TAtt, highest is the uncertainty.



\subsection{Protocol Details}
\label{sec:protocol_details}

\begin{figure}[t!]
	\centering
	\includegraphics[width=.7\columnwidth]{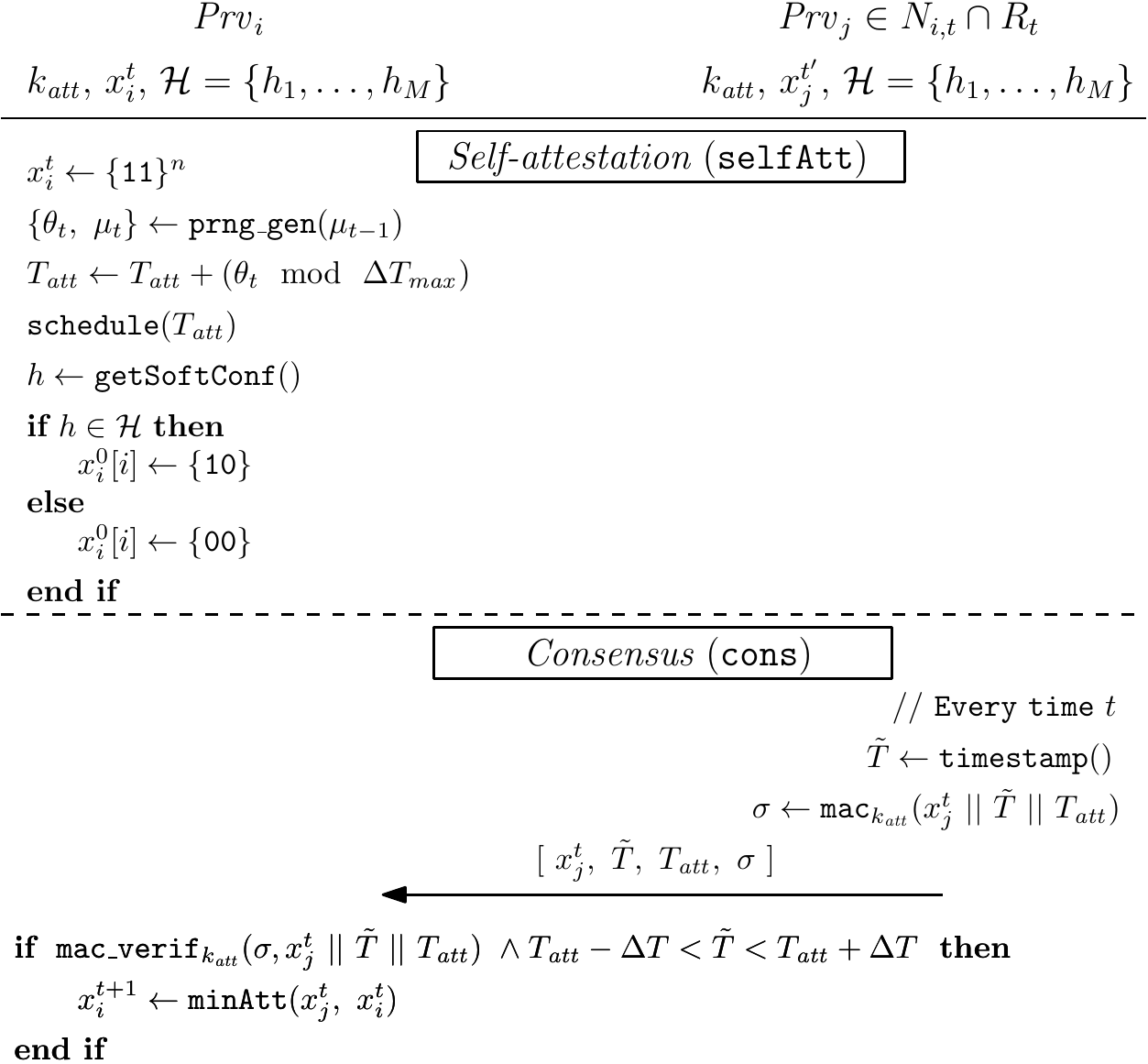}
	\caption{\protocolName's \fSelfAttest and \fConsensus procedures shown for a single reachable prover \prvi{j}.}\label{fig:protocol_details}
\end{figure}

\protocolName comprises four phases: {\em initialization}, {\em self-attestation}, {\em
	consensus}, and {\em verification} (the latter two shown in \autoref{fig:protocol_details}).

\newpar{Initialization} Similar to \cite{carpent2017lightweight}, we assume for
simplicity that nodes in the network are pre-provisioned with a shared
symmetric key \keyAtt. As discussed in \cite{carpent2017lightweight}, this is
sufficient in a software-only adversarial setting. Furthermore, we assume each
prover is provisioned with the set $\configs =
\{\confign{1},\confign{2},\dots,\confign{M}\}$ of known good configurations
(i.e., hash values); this list is either assumed to be static, or infrequently
securely updated. Finally, as in~\cite{ibrahim2017seed}, provers are
provisioned with a shared secret seed
\seedAtt that they  will use to
autonomously calculate a pseudo-random sequence of attestation times.
Alternatively, the whole sequence of attestation times may be securely
stored inside the device, at the price however, of a larger memory occupation.

\newpar{Self-attestation} The next attestation procedure is performed at a
certain point in time \TAtt, which is computed by device's secure clock using
the function \prngGen, and triggered  by secure clock's function \fSchedule.
Each prover \prvi{i} executes a self-attestation procedure. This procedure is
similar to the one in~\cite{ibrahim2017seed}, but with a major
difference: \prvi{i} measures its configuration producing a measurement value
\measurement, but instead of sending \measurement to \vrf, it performs a
``self-assessment''. More precisely, at time \TAtt each \prvi{i} computes the
hash \measurement of a predefined area of the memory --
\fGetSoftConf\footnote{The part of the memory measured during attestation can
	vary, and depends from the specific applicative scenario.}. As in
\cite{ambrosin2016sana}, this result is then checked against the (set of) good
configuration(s) \configs, pre-stored inside the device, using the \fCheck
function. The result of this operation is $\sattRes{i}\in\{0,1\}$, i.e., 1
if the configuration is a good one, and 0 otherwise. \prvi{i} further sets
$\prvBitmaskEl{i}{0}{i} = \attValOk$ if $\attRes{i}{t}=1$, and
$\prvBitmaskEl{i}{0}{i} = \attValBad$ otherwise. 
%

\newpar{Consensus} In this phase, provers corroborate their ``observations''
of the status of the network, using the distributed minimum consensus
introduced in~\autoref{sec:min_consensus}. The goal is to jointly converge to
the same ``view'' of the network, represented as a bitmask. To do that,
periodically each prover \prvi{i} broadcasts its observation, i.e.,
\prvBitmask{i}{t} at time $t$; this is the bitmask representing a snapshot of
the status of the network, together with a timestamp \timestamp, and a MAC
\signature taken over \prvBitmask{i}{t}, \timestamp, and \TAtt. Every other
prover \prvi{j} receiving a (broadcast) message, verifies the MAC \signature
-- \macVerif, checks whether \timestamp resides within a valid time interval
(to prevent reply attacks), and finally performs the minimum consensus
calculus according to \autoref{eq:min_consensus_att}. 
%


\newpar{Verification} \vrf collects the final network status \prvBitmask{}{t},
i.e., the result of the collective attestation protocol, from a prover
$\prvi{i}$, randomly chosen from \Netw. \vrf executes this final step in any moment after a
``reasonable'' amount of time \timeMax, which can be estimated by \vrf, or
simply fixed. Furthermore, the choice of $\prvi{i}$ may be dictated by
physical conditions, e.g., $\prvi{i}$ is in proximity of \vrf. The final
verification step is as simple as listening the message broadcasted by
$\prvi{i}$, verifying the MAC \signature on it (using \macVerif), and check
whether \timestamp resides within a valid time interval (i.e., within the
range $[\TAtt - \timeTolerance,~\timeMax]$). If any of the above checks
fail, \vrf outputs $\sattRes{\vrf} = 0$, i.e., the result of the collective
attestation process $\sattRes{\vrf}$ is $0$; otherwise, \vrf will output
$\sattRes{\vrf} = 1$, and the representativity value \pRepresentativity, and records all the goods and compromised devices. 

\subsection{Security of  \protocolName}\label{sec:security_analysis}

\newcommand{\Exp}{\ensuremath{\mathbf{Exp}}}
\newcommand{\ExpAdv}{\ensuremath{\mathbf{Exp}_\adv}\xspace}

We now discuss the security of \protocolName against the adversary model
introduced in~\autoref{sec:security_assumptions}. In a collective attestation
protocol, the goal of a local and/or remote adversary \adv is to modify the
configuration (e.g., firmware and/or data) of one or more provers \prv, and
evade detection from the verifier \vrf. This can be formalized as a security
experiment \ExpAdv, where \adv interacts with \prv and \vrf during the
execution of the protocol. 
After a polynomial number of steps in $\ell_\mac,
\ell_\prngGen$, \vrf outputs (\sattRes{},\pRepresentativity), where
$\sattRes{}$ is the attestation result ($\sattRes{}=1$ if it accepts attestation result, otherwise $\sattRes{}=0$) and \pRepresentativity is a measure of the quality of the fetched attestation result.
We define the result of the experiment as the output of \vrf, i.e., $\ExpAdv = (\sattRes{},\pRepresentativity)$.

We consider a collective attestation protocol for dynamic networks to be secure, if $\prob{\sattRes{} = 1 | \ExpAdv(1^\ell) = \sattRes{}}$ is negligible in $\ell = g(\ell_\mac, \ell_\prngGen)$, with $g$ polynomial in $\ell_\mac, \ell_\prngGen$. In other words, a collective attestation protocol for dynamic networks is considered secure if it is computationally infeasible for a (polynomial) attacker to induce \vrf to accept a fake attestation result. It follows that \protocolName is a secure collective attestation protocol for dynamic networks, if: (1) the MAC, e.g., a Hash-based MAC (HMAC), in use is selective forgery resistant, and (2) the adopted pseudo-random number generator is cryptographically secure.
\protocolName adopts SeED~\cite{ibrahim2017seed} to perform self-attestation at pseudo-random points in time; as such, we refer the reader to~\cite{ibrahim2017seed} for a proof of the security of self-attestation scheme, and we use it as a building block in our security discussion. 

In what follows, we will focus on the consensus part of the protocol, and discuss the security of \protocolName w.r.t. a communication adversary \advComm, a software adversary $\advSoft$, and a mobile adversary
\advMob.

\begin{compactitem}
	
	\item{\em Communication Adversary (\advComm).} 
	\advComm can try to either forge a message exchanged among provers, or the ones exchanged among a prover and \vrf, or try to ``reuse'' an old good attestation message. Both attacks will fail, since: (i) the probability for \vrf, or a prover \prv to accept a message with a forged MAC is negligible in $\ell_\mac$; and (ii) each attestation message contains both the time of attestation \TAtt, and a timestamp \timestamp, which guarantee the freshness of the received message.
	
	\item{\em Software Adversary (\advSoft).} A purely software (remote) adversary \advSoft may attempt the following attack strategies to remain undetected: (i) Modify the code responsible for attestation inside the prover; (ii) Extract \prv's security parameters; and (iii) Manipulate \prv's clock to trigger future correct measurements of current software~\cite{ibrahim2017seed}. These attacks have been proven infeasible for a polynomial attacker in~\cite{ibrahim2017seed}. Furthermore, similarly to self-attestation, consensus operations are performed by a piece of code protected by secure boot.
	
	\item{\em Mobile Software Adversary (\advMob)} Assume at time $t$ a mobile (software-only) adversary \advMob compromises a prover \prvi{j} before the next attestation time \TAtt, and tries to avoid detection by simply not taking part to the protocol, e.g., physically move out of the communication range of other provers in \Netw. 
	As a result, no prover in \Netw will modify the part of the bitmask related to the compromised prover, \prvBitmaskEl{j}{t}{j}, whose value will remain \attValUnknown; this information will not affect the decision of \vrf, who may decide, adopting a conservative approach, to consider \prvi{j} as compromised.
	
\end{compactitem}


\section{Implementation and Evaluation}
\label{sec:pads_evaluation}



\begin{figure}[t!]
	
	\centering
	\includegraphics[width=.65\columnwidth]{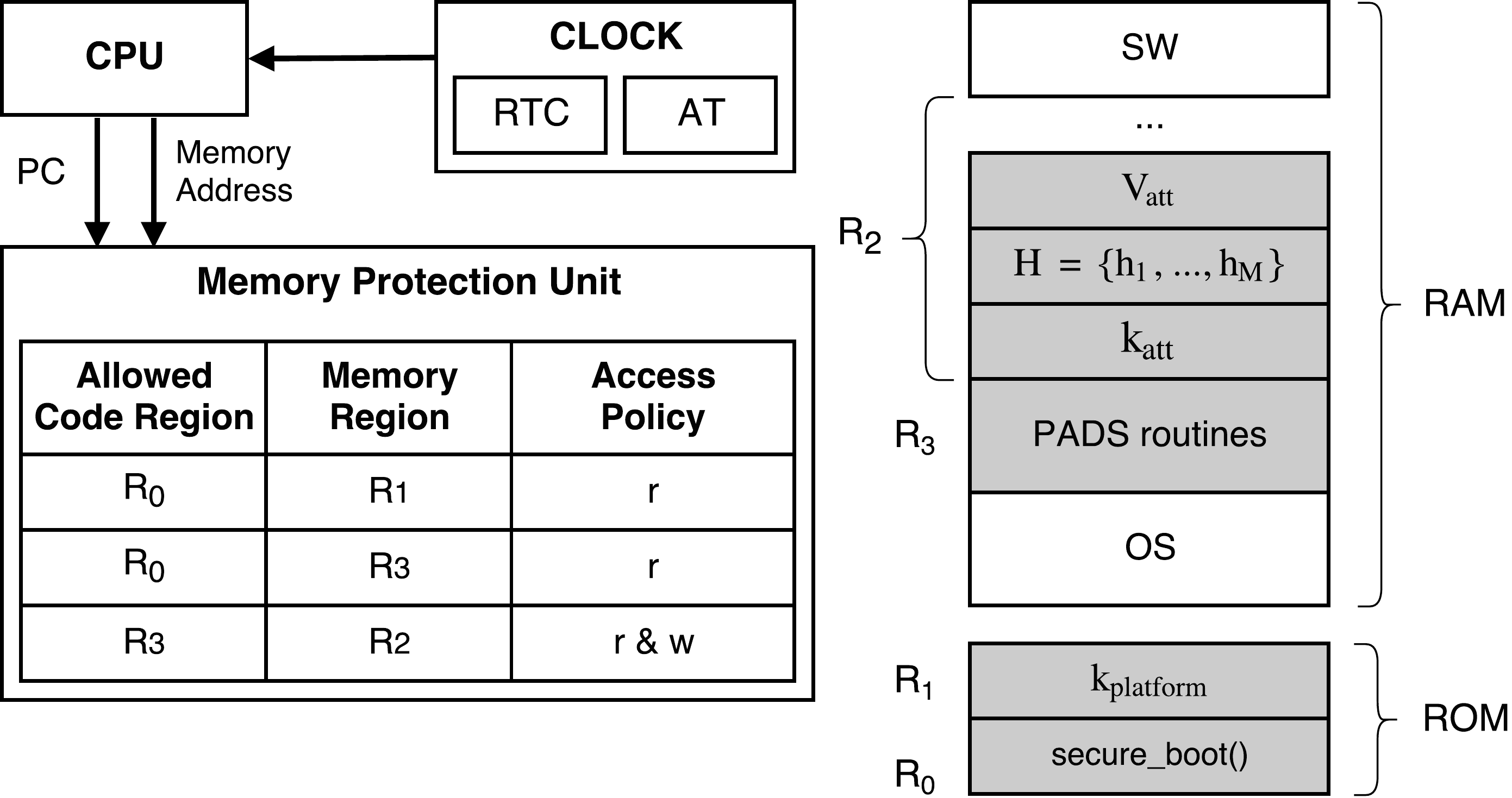}
	\caption{Implementation of \protocolName based on SeED~\cite{ibrahim2017seed} and TrustLite~\cite{koeberl2014trustlite}.}\label{fig:PADS_based_on_SeED}
	
\end{figure}

We base our evaluation of \protocolName on SeED~\cite{ibrahim2017seed}, a recently proposed protocol for devices self-attestation. \autoref{fig:PADS_based_on_SeED} shows the implementation design of \protocolName on top of the enhanced version of
TrustLite~\cite{koeberl2014trustlite} used in~\cite{ibrahim2017seed}.
SeED~\cite{ibrahim2017seed} extends TrustLite with a Real-Time Clock (RTC) that is
not modifiable via software (i.e., that is write protected), and an
Attestation Trigger (AT) that updates and monitors the value of a timer stored
in a secure register.

We quantify \protocolName performance based on runtime, energy consumption,
and memory overhead. 
Furthermore, we evaluate the ability for \protocolName to
``cover'' a sufficiently wide area of the ``reachable'' provers population, within a certain time $t$, using the following notion of {\em coverage}: 

\begin{definition}[Coverage]
	We say that at time $t$ \protocolName has coverage $\coverage{X}=Y$, if at
	least a portion $X\in [0,1] $ of the provers in \Netw hold information of at
	least a portion $Y\in [0,1] $ of the reachable provers population.
\end{definition}

As an example, in a network of 100 provers, where 10\% of them are unreachable,
\coverage{80} = 90\% means that 80\% of the 90 reachable provers in \Netw hold
information about the 90\% of them.
Intuitively, the higher the level of coverage, the higher the number of update
steps that are required. This has clearly an impact in the performance of the
protocol, that we estimate and present in the following. As in~\cite{asokan2015seda}, in our evaluation we consider a SHA-1 based HMAC.

\newpar{Memory Overhead} In a shared key setting, i.e., all
the provers share the same symmetric key \keyAtt for attestation, the required memory
overhead for \protocolName depends on $\ell^{\keyAtt}$, the number of  allowed
configurations in \configs, and the size of the bitmask, which grows linearly
with the number of provers \size. Overall, let $M = |\configs|$, and
considering each element of \configs~of 20~B, regardless from the coverage, we
can quantify the memory overhead of \protocolName as $\ell^{\keyAtt} + 2\times n + 160*|\configs|$ bits.

\newpar{Communication Overhead} During \protocolName's consensus phase, each
prover \prvi{i} transmits (and/or receives), at every time $t$: its bitmask
\prvBitmask{i}{t} ($2\times n$ bits), the attestation time \TAtt (32 bits) ,
the timestamp \timestamp (32 bits), and one HMAC (160 bits). In total: $2\times
n + 224~\text{bits}$. Clearly, depending on the underlying layer 2 protocol
and the size of the network, we may have more or less fragmentation of the
transmitted data. In low-power settings, it is desirable to limit the number
of transmitted frames, and thus, the ``maximum'' size of the network that
\protocolName can serve depends on the payload size offered by the underlying layer 2
protocol. 

%
%

\newpar{Energy Consumption} 
We base our estimation of the overall energy consumption on the required
energy for sending and receiving a single  \protocolName message, and to
compute the main cryptographic operations. 
Let \Esend, \Erecv, \Emac, \Emin and \Eatt denote, respectively, the energy
required to send a byte, receive a byte, calculate one HMAC, calculate the
minimum consensus over the received bitmask, and perform self-attestation.
Recall that, at each round $t$ each prover \prvi{i} sends: the bitmask
\prvBitmask{i}{t}, the attestation time \TAtt, a timestamp \timestamp, and one
HMAC. Thus, we can estimate the energy consumption for sending and receiving a single
\protocolName message for \prvi{i} as:
\begin{align}
	\Esend^{\prvi{i}} \leq \Esend\times(28 + \frac{2\times n}{8}) \nonumber;~~~~
	%
	\Erecv^{\prvi{i}} \leq \Erecv\times(28 + \frac{2\times n}{8}).\nonumber
\end{align}

Let $m$ be the number of rounds of consensus required by the protocol. To
provide an overall estimate of the energy required by \protocolName, without
loss of generality we assume each prover shares its bitmask at fixed time
intervals $t_1, t_2, \dots, t_m$. Recall that \Neigh{i}{t} is the set of
neighbors of prover \prvi{i} at time $t$. At each time $t$, a prover \prvi{i}
computes an HMAC, broadcasts a packet, receives a number of broadcast messages
from its neighbors \Neigh{i}{t}, and verifies the HMAC associated with them.
We estimate the energy required by \protocolName for a prover
\prvi{i} as:
\begin{align}
	\Epads^{\prvi{i}} 	&\leq \TAtt + \nonumber \sum_{t = t_1}^{t_m}{\Emac + \Esend^{\prvi{i}} + (\Emac + \Erecv^{\prvi{i}}) *|\Neigh{i}{t}\cap\responsive{t}|}. \nonumber
\end{align}



\newpar{Runtime}  Similar to previous works, we evaluated \protocolName's
runtime using Omnet++~\cite{omnet}, and the MiXiM~\cite{mixim} framework (for
realistic communication protocol simulation and mobility). We considered
networks of medium-large sizes, from 128 to 16,384. In our simulation we
consider provers to have specifications comparable to the one
in~\cite{ibrahim2017seed}, and thus used their reported micro-benchmarks as
parameters of our simulation. All the communications are carried out over the
IEEE 802.15.4 MAC layer protocol, the de-facto standard protocol for
IoT~\cite{asokan2015seda,ambrosin2016sana,ambrosin2016despicable}. The IEEE
802.15.4 protocol provides a maximum data rate of 250~Kbps, a maximum coverage
of 75~m, and a frame size of 127~B.  We investigated the performance of
\protocolName in two scenarios: (1) A scenario where provers move following a random path. To model provers
mobility, we randomly deployed provers over a simulated area of size
proportional to the number of devices (e.g., simulating an area covered by a
swarm of drones), starting from $1,000\times1,000$~m$^2$ for 128 provers. The
random movement of provers makes the network dynamic and loosely connected. (2) A static scenario, as considered in previous work, where provers are
organized in three topologies of various branching factor; in this setting, we
compare the runtime of \protocolName w.r.t. SANA~\cite{ambrosin2016sana}.

Before presenting our results, in order to measure the runtime of
\protocolName we define the notion of Minimum Coverage Time (MCT):

\begin{definition}[Min Coverage Time]
	The Minimum Coverage Time (MCT) for \protocolName, given a desired coverage
	level $\coverage{X} = Y$, is the minimum amount of time $t$ necessary to reach
	\coverage{X}. Formally: $\arg\min_{t} \coverage{X} = Y.$
\end{definition}

We measured the runtime of \protocolName as the average MCT. We used delays to simulate provers' internal operations, according to~\cite{asokan2015seda} and~\cite{ibrahim2017seed}. We considered 48~ms
as the time it takes to compute both a hash, and an HMAC~\cite{asokan2015seda}.
Furthermore, for every generated attestation response, the overhead introduced
by the self attestation part of \protocolName is
187~ms~\cite{ibrahim2017seed}. This is approximated by the generation of 32
random bits, and the calculation of a hash over the amount of memory to be attested. Note that, different from~\cite{ibrahim2017seed}, we do not compute
an HMAC on the measurement, as the measurement is not sent to the verifier,
but checked locally; thus, our evaluation is conservative.

\begin{figure}[t]
	\centering
	%
	\subfloat[\protocolName runtime]{
		\includegraphics[width=.48\textwidth]{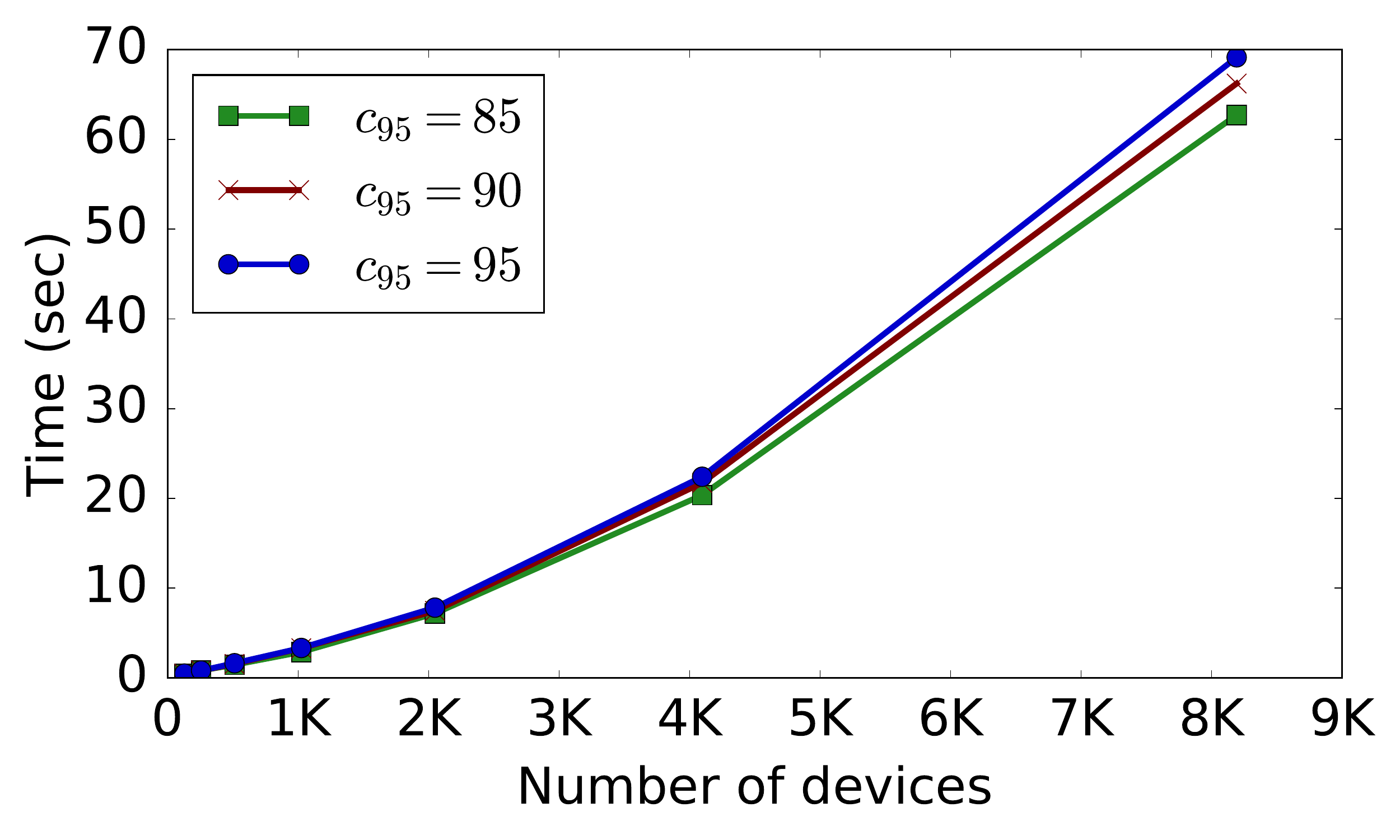}\label{img:runtime}}
	\subfloat[Variation of \coverage{95} and avg. steps number for $\size = 8196$]{
		\includegraphics[width=.48\textwidth]{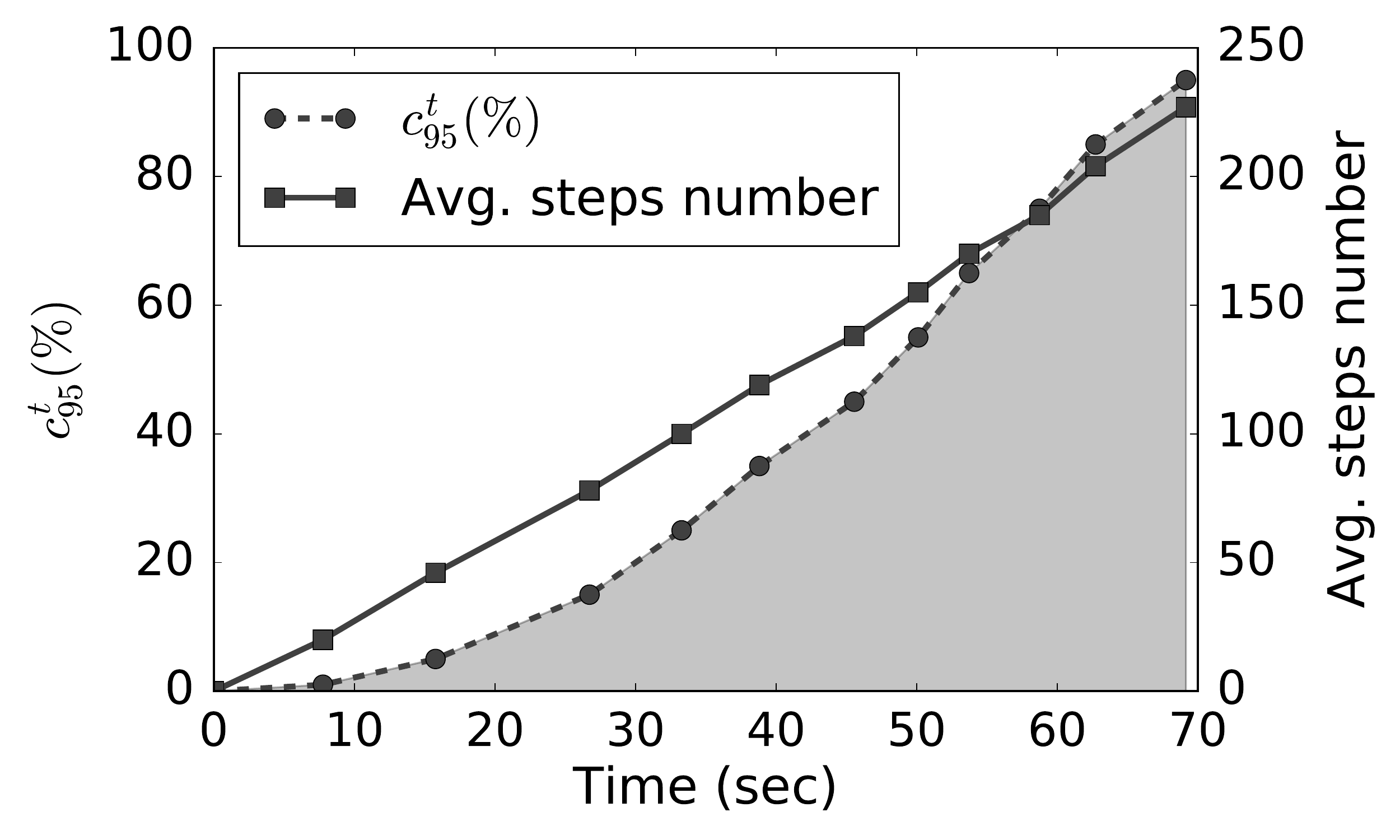}\label{img:cov_and_steps_variation}}
	\caption{Evaluation of \protocolName in mobile wireless setting, varying number of provers and size of the area (proportionally to the number of provers); broadcast frequency is 500~ms. \autoref{img:runtime} shows the runtime of \protocolName considering different values of \coverage{95}; \autoref{img:cov_and_steps_variation} shows the relation between \coverage{95} and average number of steps.}
	\label{fig:runtime_per_coverage}
\end{figure}

For different levels of coverage, we measured the average time for that
coverage to be reached (average of 50 runs), considering network of small-large sizes, from 128 to 8,196 devices; we adopted a static rate for broadcasting the bitmask, i.e., 500~ms. Furthermore, we considered all the provers, either good or compromised, to participate to the attestation process. Results are presented in
\autoref{fig:runtime_per_coverage}. 
As we can see from \autoref{img:runtime}, \protocolName can reach,
on average, 95\% of coverage for the 95\% of the population (i.e.,
$\coverage{95}=95$) with a MCT lower than 70 seconds, for populations of
8,196 devices. Furthermore, as indicated in \autoref{img:cov_and_steps_variation}, the coverage grows more than linearly over time (shown for \coverage{95} and $\size = 8196$ in \autoref{img:cov_and_steps_variation}).

In a completely dynamic setting,
\protocolName runtime shows a non-negligible growth. This is mainly due not
only to the increase of the population, but also on the amount of data to be
transmitted (which grows linearly in the number of provers). Despite this
growth, \protocolName presents a remarkably manageable overhead for large
networks, which makes it a good match for practical applications. Furthermore, we stress that, while previous work would have failed completing the attestation process due to the high degree of mobility of this scenario, \protocolName allows to complete the attestation process.



\begin{figure}[t]
	\centering
	\subfloat[$\mathit{br}=2$]{
		\includegraphics[width=.48\columnwidth]{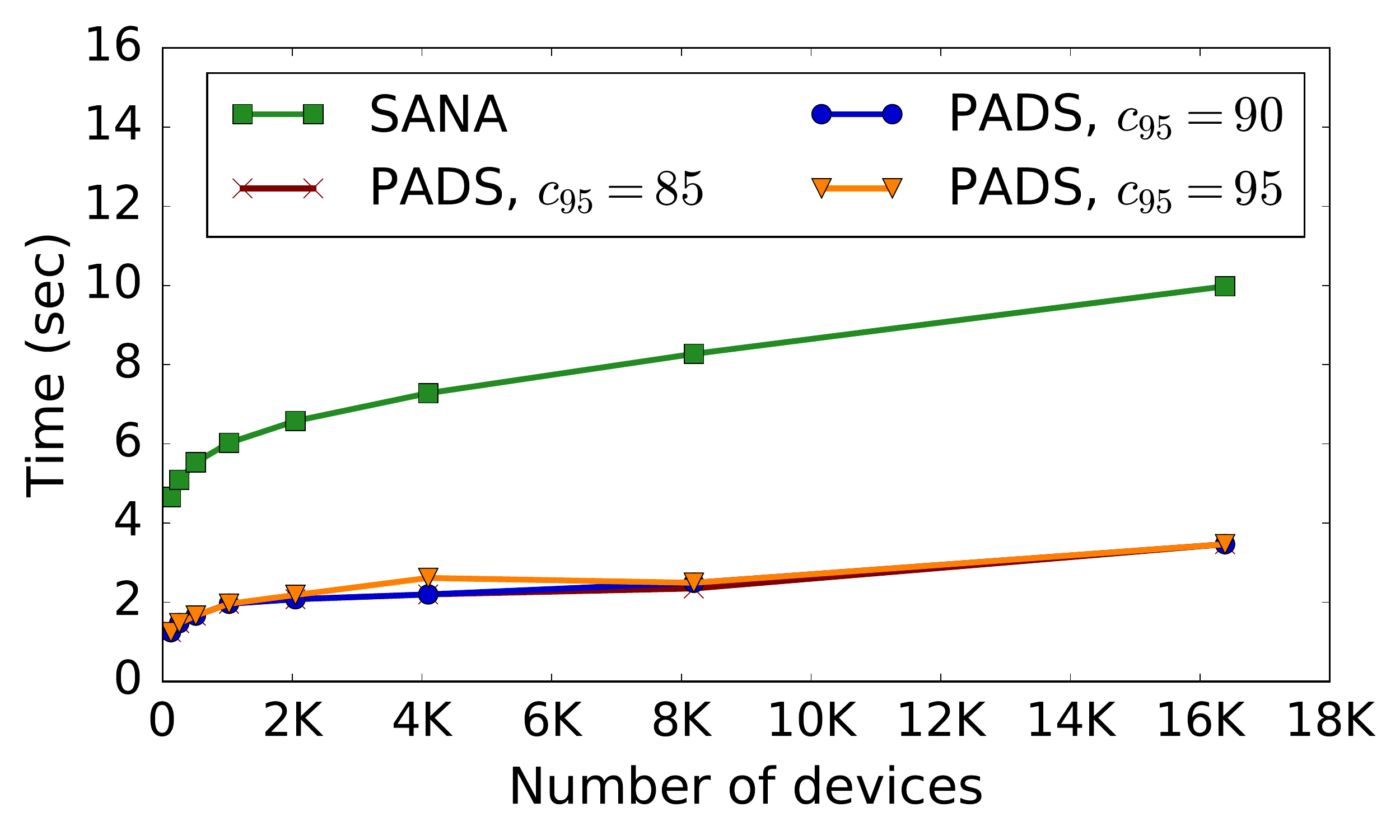}}\label{sana_vs_pads_tree_brf2_X95}
	\subfloat[$\mathit{br}=3$]{
		\includegraphics[width=.48\columnwidth]{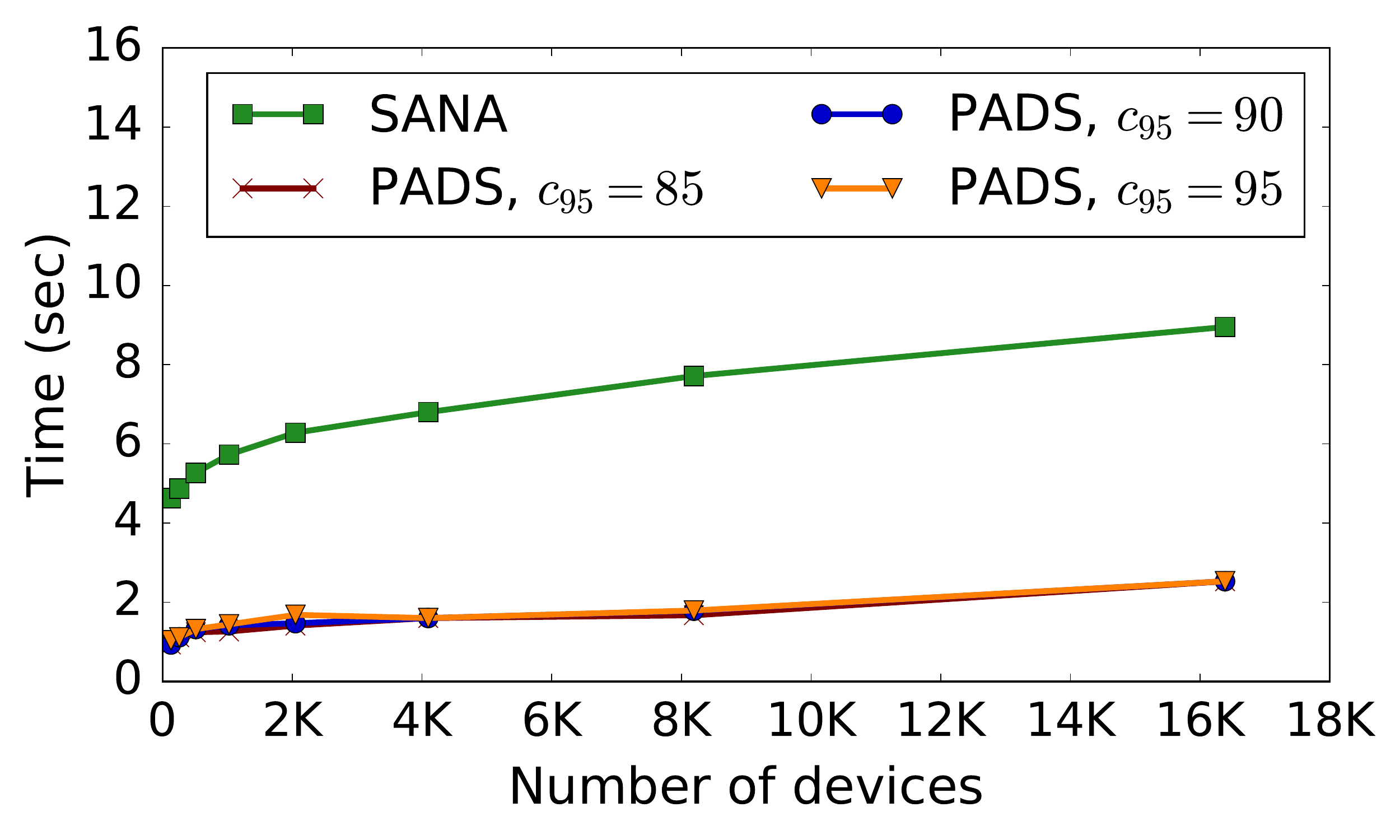}   }\label{sana_vs_pads_tree_brf3_X95}
	
	\caption{Runtime of \protocolName vs SANA~\cite{ambrosin2016sana}, varying number of nodes and \coverage{X}, with $X = 95\%$. We considered up to 50 different configurations, and 60\% compromised provers.}
	
	\label{fig:runtime_sana_vs_pads}
\end{figure}

We further compare \protocolName with SANA~\cite{ambrosin2016sana}, a recently proposed collective attestation scheme, which outperformed previous work in the literature. We run both protocols on static tree topologies of branching factor ($\mathit{br}$) 2 and 3, using for \protocolName a broadcast frequency of 100~ms. Results are shown in~\autoref{fig:runtime_sana_vs_pads}. As we can see, \protocolName has a lower runtime (i.e., MTC) compared to SANA, mainly due to the more lightweight cryptography involved. Furthermore, we can see how the runtime of \protocolName diminishes with the branching factor of the tree topology. This confirms the low overhead of our protocol, even for large networks of more than 16,000 devices.
We finally remind that the time necessary for verifier to obtain the final attestation results in \protocolName is the time necessary for querying a single device; that is, the collection of the final result is separate from the collection and agregation of the single attestation proofs. In SANA, instead, the attestation protocol starts after the verifier queries the first node; in this case, \vrf has to wait until the protocol is concluded.

%


\section{Conclusions}\label{sec:conclusions}



This paper presented \protocolName, an efficient, practical, and secure
protocol for attesting potentially large and highly dynamic networks of
resource-constrained autonomous devices. \protocolName overcomes the
limitations of previous works in the literature on collective attestation; it
uses the recently proposed concept of self attestation, and turns the
collective attestation problem into a minimum consensus one. We showed the
performance of \protocolName via realistic simulations, in terms of devices
capabilities and communication protocol, confirming both the practicality and
efficiency of \protocolName.

As a future work, we will investigate ways to reduce the complexity of the
protocol in terms of both communication and required processing for devices;
we will explore the use of Bloom Filters as a way to reduce the size of the
payload, as well as adopting more intelligent techniques to selectively pick
messages to use for consensus;
we will modify the protocol to manage nodes joining (or leaving) the network after the initial setup.

\bibliographystyle{abbrv}
\bibliography{IEEEabrv,bibliography}

\appendix


\section{Discussion}\label{sec:summary_and_discussion}


\newpar{Advantages} The first and probably most important advantage that \protocolName has w.r.t. previous works, is that is suitable for highly dynamic, mobile unstructured
topologies, as confirmed by our evaluation. The use of consensus makes the
whole proliferation of the attestation results resilient to both temporary
device disconnections, and topology changes. Most of the previous works in the
area were designed for static topologies, over which a spanning tree
should be maintained, with non-negligible overhead; as such, they would most
certainly fail in this scenario~\cite{asokan2015seda,ambrosin2016sana,ibrahim2016darpa,carpent2017lightweight,ibrahim2017seed}; one exception is the work
in~\cite{kohnhauser2017scapi}, which provides adaptability to changes in
the topology, at the price, however, of a complex topology maintenance.


The second main advantage of \protocolName, is the capability of the verifier
to obtain the status of the population by simply contacting a random prover.
This, again, makes \protocolName very resilient against node failures or
sudden changes in the topology. This is due to the fact that the whole status
of the network is shared among every prover; furthermore, a verifier has the
ability to query a device for the status of the network at {\em any point in
time}, being assured that this would return a valid attestation result, with a
utility (i.e., a coverage) dependent on the status of the protocol.

Finally, while growing more than linearly in the number of devices (mostly because the amount of information that each prover must exchange grows linearly in the number of nodes in the system), the overhead introduced by \protocolName is manageable in resource-constrained environments, as showed by our simulations in~\autoref{sec:pads_evaluation}; furthermore, in the same (static) setting as previous work, \protocolName shows comparable if not superior performance w.r.t. the state of the art. This makes \protocolName a good candidate for attestation in scenarios where other exact methods would fail, e.g., swarms of drones for surveillance. We stress that \protocolName for the first time {\em enables} swarm attestation in dynamic topologies of autonomous devices, while previous protocols works only on static networks. Therefore, by design, \protocolName trades scalability for adaptability and resiliency in dynamic scenarios.

\newpar{Limitations} While \protocolName brings numerous advantages w.r.t. the state of the art, we are conscious that network dynamics is not network security's friend, and we acknowledge that \protocolName is not perfect. 



First, the final consensus is a representation of the status of the devices composing the network at the time the last attestation has been triggered by the device's secure clock. Therefore, a device compromised after such timestamp can remain undetected\footnote{Note that,~\cite{ibrahim2017seed} uses pseudo-random generation of attestation times to tackle mobile attackers compromising a device and then re-establish a correct configuration between consecutive attestation runs.}. Unfortunately, this is a well known limitation of most attestation protocols, commonly defined as ``time of check to time of use''~\cite{steiner2016attestation}. 

Finally \protocolName aligns with previous works that considered a software-only attacker~\cite{asokan2015seda,ibrahim2017seed,carpent2017lightweight}, and thus a single physically compromised device is sufficient to reveal the secret shared among all other devices, enabling the attacker to inject completely fake attestation reports in the network. However, even in the simplest case, we argue that \protocolName shows some degree of resiliency with respect to previous works that made the same assumptions. Intuitively, this is provided by the combination of the following two properties of \protocolName: (a) in a generic setting, information is not processed along a tree; (b) data combination uses a minimum consensus approach. Property (a) suggests that there will be a certain degree of redundancy in the propagated information, which every (honest) device will receive from a non-constant neighborhood of (honest and malicious) devices; furthermore, \vrf will collect the final result from one or more random provers. In this case, the dynamic nature of the network topology is a friend of security. Consider for example the protocol in~\cite{asokan2015seda}: here, a single physically compromised device at a specific position in the spanning tree could ``fake'' the attestation report of a whole subtree of devices, even if the devices in the tree are only software compromised. In \protocolName, instead, in order to make sure a whole group of devices can evade detection, the attacker will have to physically compromise all of them (to be able to have access to the shared key, and craft a bitmask for each). Indeed, due to Property (b), it is unfeasible even for a powerful attacker with access to the shared secret, to change the value of a software-compromised device from \attValBad to \attValOk, if previously ``propagated'' in the network by honest provers as \attValBad. This requires a much higher effort for the attacker compared to~\cite{asokan2015seda}.
Physically compromised devices have been considered in~\cite{ambrosin2016sana} and \cite{ibrahim2016darpa}. However, as previously stated, both works are unsuitable for dynamic topologies; furthermore,~\cite{ambrosin2016sana} is quite costly in terms of required computation (see~\autoref{sec:pads_evaluation}).

\end{document}